\documentclass[showpacs,amsmath,amssymb,superscriptaddress,footinbib]{revtex4}

\usepackage{graphicx}
\usepackage{epsfig}
\usepackage{hyperref}
\usepackage{dcolumn}
\usepackage{bm}
\usepackage{epstopdf}
\usepackage{verbatim} 

\begin{document}

\author{F. K. Diakonos}
\email{fdiakono@phys.uoa.gr}
\affiliation{Department of Physics, University of Athens, GR-15771 Athens, Greece}
\author{A. K. Karlis}
\email{A.Karlis@warwick.ac.uk}
\affiliation{Department of Physics, University of Warwick, Coventry CV4 7AL, United Kingdom}
\affiliation{Department of Physics, University of Athens, GR-15771 Athens, Greece}
\author{P. Schmelcher}
\email{pschmelc@physnet.uni-hamburg.de}
\affiliation{Zentrum f\"ur Optische Quantentechnologien, Universit\"{a}t Hamburg, Luruper Chaussee 149, 22761 Hamburg, Germany}
\affiliation{The Hamburg Centre for Ultrafast Imaging, Luruper Chaussee 149, 22761 Hamburg}

\date{\today}

\title{Intermittency induced long-range cross-correlations}

\begin{abstract}
Cross-correlations are usually considered to emerge through interaction between particles. Here we present a mechanism capable to generate power-law cross-correlations between non-interacting particles exposed to an external potential. This phenomenon can occur as an ensemble property when the external potential induces intermittent dynamics, providing laminar and stochastic phases of motion. We have strong indications that the divergence of the mean residence time in the laminar phase of the single particle motion - sporadic dynamics - is a sufficient condition for the emergence of long-range cross-correlations. We argue that this mechanism may be relevant for the occurrence of collective behaviour in critical systems.
\end{abstract}

\maketitle

Complex systems usually consist of several dynamical components interacting in a non-linear fashion. Cross-correlations are then used in order to explore the inter-dependence in the time evolution of these components measured in terms of specific quantities characterizing each component. In this context, the existence of cross-correlations has been demonstrated in a wide class of dynamical systems ranging from nano-devices \cite{Sam2003} to atmospheric geophysics \cite{Yam2008}, seismology \cite{Cam2003}, finance \cite{Lal1999,Tumm2005,Stan2008,Taka2010,Stan2010}, physiology and genomics \cite{Stan2010}. Of special interest is the case of long-range (power-law) cross-correlations (LRCC) which, being scale free, may be associated with the appearance of characteristics of criticality in the dynamics of the considered complex system. Such a behaviour has been observed, among other examples, in price fluctuations of the New York Stock Exchange during crisis \cite{Stan2010}, physiological timeseries of the Physiology Sleep Heart Health Study (SHHS) database \cite{Stan2010}, the spatial sequence describing binding probability of DNA-binding proteins to genes at different locations on mouse chromosome 2 \cite{Stan2010} and in flocks of birds \cite{Cavag2010}. All these findings indicate that the presence of power-law cross-correlations is a quite general property of the dynamics of complex systems. Even more, very recently geometry induced power-law cross-correlations have been also observed in a coarse-grained description of the dynamics of an ensemble of non-interacting particles propagating in a Lorentz channel \cite{Kar2012}. This clearly pauses the question of the origin and mechanisms of cross-correlations in particle systems.

Up to now the theoretical treatment of cross-correlations is based on statistical approaches and their microscopic origin is to a large extent unclear. In the present Letter we identify the dynamical mechanisms leading to LRCC and show specifically that intermittent dynamics obeyed by each component separately generates LRCC between the different components even if they do not interact with each other. In addition we provide strong evidence that a sufficient condition for the emergence of such scale-free LRCC is the divergence of the mean length of the laminar phase in the intermittent dynamics of each component. 

The prototype model we will use to demonstrate our arguments is a system of $N$ non-interacting particles each with an one-dimensional phase space determined by the variable $x^{(i)}$ ($i=1, 2, ..., N$). We do not further specify $x^{(i)}$: in a real system it can be for example the position or the momentum of particle $i$ or any other property characterizing its state (partially or totally). We use a version of the well-known Pomeau-Manneville map of the interval \cite{Pom1980}, consisting a normal form of intermittent dynamics, to describe the time evolution of $x^{(i)}$ of each particle independently:
\begin{equation}
x^{(i)}_{n+1}=\left\lbrace \begin{array}{c} x^{(i)}_n + u^{(i)} ( x^{(i)}_n )^{z^{(i)}}~~~;~~~x^{(i)}_n \in (0,x^{(i),*}) \\  r^{(i)}_n x^{(i),*}~~~~~~~~~~~~~~~~~~~~~;~~~x^{(i)}_n \in ( x^{(i),*}, 1] \end{array} \right. 
\label{eq:eq1}
\end{equation}
for $i=1,2,...,N$. In Eq.(\ref{eq:eq1}) $u^{(i)}$ are positive constants, $z^{(i)}$ are characteristic exponents fulfilling $z^{(i)} > 1$ and $r^{(i)}_n$ are random numbers uniformly distributed in $(0,1]$. The quantity $x^{(i),*}$ represents the upper boarder of the phase space region ($(0,x^{(i),*})$) within which the evolution of the particle dynamics is laminar. Notice that in 
Eq.~(\ref{eq:eq1}) there is no coupling term between phase space variables of different particles since there is no mutual interaction. This simple model is quite general capturing all the basic dynamical ingredients necessary for the development of cross-correlations as we will see in the following. To avoid unnecessary complexity we further simplify the model assuming: $u^{(i)}=u$ and $z^{(i)}=z$ for all $i$. Note that the end of the laminar region $x^*$ is not strictly defined. One possible choice, which we use in the following, is to fix $x^*$ as the pre-image of $1$, i.e. as the solution of the equation $x^* + u (x^*)^z = 1$. A second choice is to set it equal to $\tilde{x}^*=\left( \frac{1}{u z} \right)^{\frac{1}{z-1}}$ being the $x$-value for which the non-linear term in Eq.~(\ref{eq:eq1}) becomes equal in magnitude with the linear term. These two values ($x^*$ and $\tilde{x}^*$) are very close to each other for almost all values of $z$ and our results for the cross-correlations, shown below, do not depend on this choice. Using equation (\ref{eq:eq1}) we evolve the considered particle system in discrete time. Different particles correspond to different trajectories i.e. trajectories starting from a different initial condition. Thus we propagate a set of $N$ trajectories. At each point of a trajectory we define an observable quantity $A^{(i)}_n \equiv A(x^{(i)}_n)$ (which can be $x^{(i)}_n$ itself).
Then the cross-correlation function with respect to the observable $A$ is defined as:
\begin{equation}
CC_A(m)=\frac{2}{N(N-1)}{\displaystyle{\sum_{i < j}}} \frac{\langle A^{(i)}_{n} A^{(j)}_{n+m} \rangle - 
\langle A^{(i)}_n \rangle \langle A^{(j)}_{n+m} \rangle}{\sigma_A^{(i)} \sigma_A^{(j)}}
\label{eq:eq2}
\end{equation}
where $\langle ... \rangle$ denotes time averaging while $\sigma_A^{(i)}$ and $\sigma_A^{(j)}$ are the standard deviations of $A^{(i)}$ and $A^{(j)}$ respectively.

A typical characteristic of the intermittent dynamics is that for any trajectory the $x$-values in the laminar region are very close to the diagonal $x^{(i)}_{n+1}=x^{(i)}_n$ since the increase $\Delta x^{(i)}_n = x^{(i)}_{n+1} - x^{(i)}_{n}$ of $x^{(i)}_n$ there is very slow. It is important to notice that the cross-correlations, calculated using Eq.~(\ref{eq:eq2}), appear in the phase space variable $x$ while the increments $\Delta x$ of different trajectories are clearly uncorrelated. This behaviour is illustrated in Figs.~1a,b where we plot $CC_x(m)$ (Fig.~1a) and $CC_{\Delta x}(m)$ (Fig.~1b) for an ensemble of trajectories evolving according to the law given in Eq.~(\ref{eq:eq1}) with $u=1$ and $z=2.5$. 

\begin{figure}[htbp]
\includegraphics[width=8.5 cm]{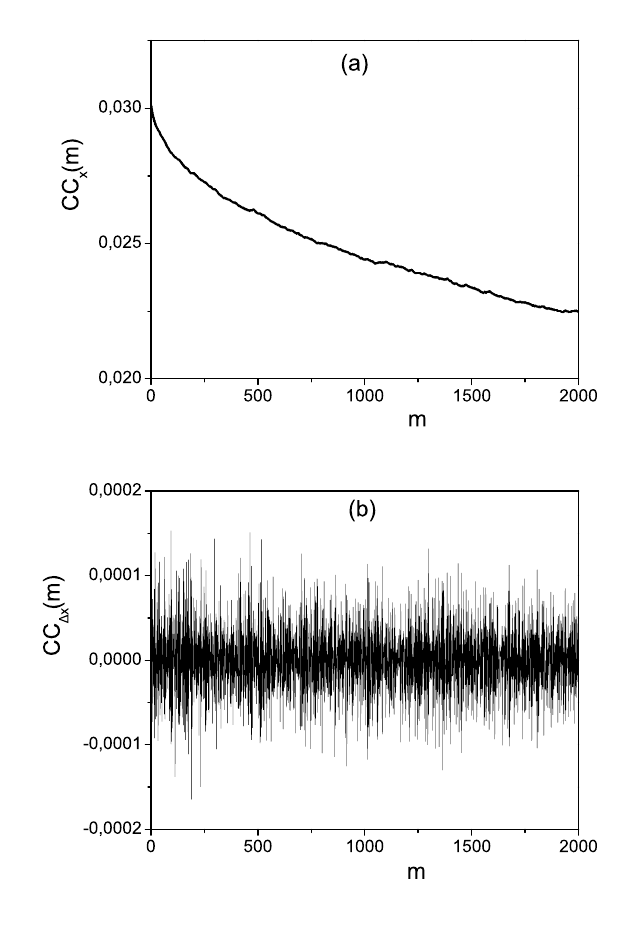}
\caption{\label{fig1} (color online). (a) The cross-correlation function $CC_x(m)$ as defined by Eq.~(\ref{eq:eq2}) using $10^4$ trajectories each of length $10^5$ and, (b) $CC_{\Delta x}(m)$ for the same ensemble of trajectories. The parameters of the intermittent dynamics are $z=2.5$ and $u=1$.}
\end{figure}

To illuminate further the origin of the cross-correlations shown in Fig.~1a we explore their dependence on the value of the parameter $z$. For this task we calculate $CC_x(m)$ for various values of $z$ using ensembles of $10^4$ trajectories with length $10^5$ for each case. For $z > 2$ we find an algebraic decay of $CC_x(m)$ with increasing $m$, having an exponent which depends on $z$, while for $z <2$ $CC_x(m)$ fluctuates around zero indicating the absence of cross-correlations in this case. The behaviour of $CC_x(m)$ for $z < 2$ is demonstrated in Fig.~2 where we plot as a typical example the result obtained for $z=3/2$. It is worth to mention here that a distinction between the properties of intermittent dynamics for $z < 2$ and $z > 2$ has been already discussed in \cite{Gasp1988} where the term {\it sporadicity} is introduced for the description of the $z > 2$ case.

\begin{figure}[htbp]
\includegraphics[width=8.5 cm]{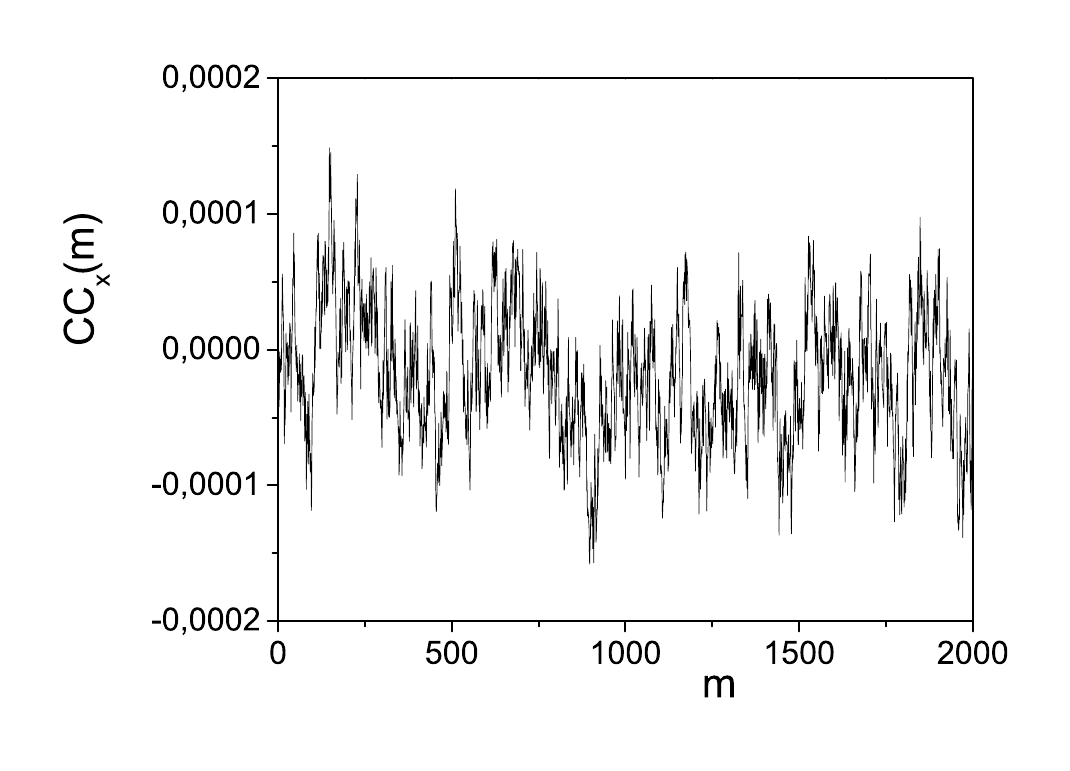}
\caption{\label{fig2} (color online). The cross-correlation function $CC_x(m)$ for the intermittent map of Eq.~(\ref{eq:eq1}) with parameters $z=1.5$ and $u=1$. For the numerical simulations we used an ensemble $10^4$ trajectories each of length $10^5$. }
\end{figure}

In order to explain the different behaviour of the cross-correlation functions for $z > 2$ and $z < 2$ we consider the distribution of the laminar phase lengths or as it is often also named the distribution of the waiting times in the laminar region. It is well known that this distribution obeys asymptotically ($\ell \gg 1$) a power-law of the form $\rho(\ell) \sim \ell^{-\frac{z}{z-1}}$ \cite{Schuster,Proc1983} where $\ell$ is the laminar phase length. For $z > 2$ the mean laminar length $\langle \ell \rangle$ diverges while for $z < 2$ it is finite \footnote{We refer here to the divergence implied by the asymptotic behaviour for $\ell \to \infty$. In the small $\ell$ region there is always a natural cut-off since $\ell \geq 1$}. Thus the divergence of $\langle \ell \rangle$ should be related with the emergence of cross-correlations between the particles. This in fact can be explained as follows: if $\langle \ell \rangle$ is infinite then the conditional probability that the particle $j$ at an instance $n+m$ is in the laminar region provided that the particle $i$ was in the laminar region at instance $n$ is finite and decays algebraically with increasing $m$. 

To understand better this behaviour it is useful to develop a symbolic code for the intermittent dynamics in Eq.~(\ref{eq:eq1}). Such a symbolic representation of the dynamics of the Pomeau-Manneville intermittent map capturing several details like the re-injection rate in the laminar region (and therefore the invariant density in the immediate neighbourhood of the marginally unstable fixed point) is proposed in \cite{Gasp1988}. Here we are interested mainly to isolate the dynamical properties leading to the emergence of cross-correlations avoiding the influence of other detailed aspects of the intermittent dynamics. Therefore we will use a much simpler code, mapping $x$ in the laminar region ($x \in [0,x^*]$) to $0$ and  $x$ out of the laminar region ($x \in (x^*,1]$) to $1$. Such a code is used in \cite{Proc1983} to calculate power-spectra of intermittent systems. In practice we use the full dynamics of Eq.~(\ref{eq:eq1}) to generate the ensemble of intermittent trajectories and then we replace the $x$-values in each time-series by $0$ or $1$ according to the previously described rule. Subsequently we calculate the cross-correlation function $CC_s(m)$ for different $z$-values using the binary sequences generated by the symbolic dynamics from the trajectories of the map in Eq.~(\ref{eq:eq1}). 

Furthermore, in order to get rid of any other dynamical effects which may still be present in the symbolic dynamics, we introduce a stochastic model containing only the information of the laminar length distribution to simulate the emergence of cross-correlations. We assume a process consisting of two phases defined as follows: (i) a stochastic variable $\xi$ takes the value 1 in the irregular phase and the value 0 in the laminar phase and (ii) the length of the irregular phase is always 1  while the laminar length probability density is a power-law with exponent $-z/(z-1)$ ($z$ being the exponent in the intermittent map of Eq.~(\ref{eq:eq1})). Then we generate an ensemble of realizations of this process and calculate the cross-correlation function $CC_r(m)$ for this ensemble. Despite the simple form of the map providing the intermittent dynamics, large scale computational efforts ($10^5$ trajectories have been propagated for $10^6$ iterations) are needed to achieve convergence of the long-time behaviour of the cross-correlation function. In Fig.~3 we show the results obtained for $CC_A(m)$ with $A=s,r$ for $z=2.5,3,4,5$. The coloured triangles correspond to $A=s$ while the red lines to $A=r$. We observe a very good agreement between the two results for each $z$ value. This is a strong indication that the quantity determining the properties of the cross-correlation function is indeed the laminar length distribution.

\begin{figure}[htbp]
\includegraphics[width=8.5 cm]{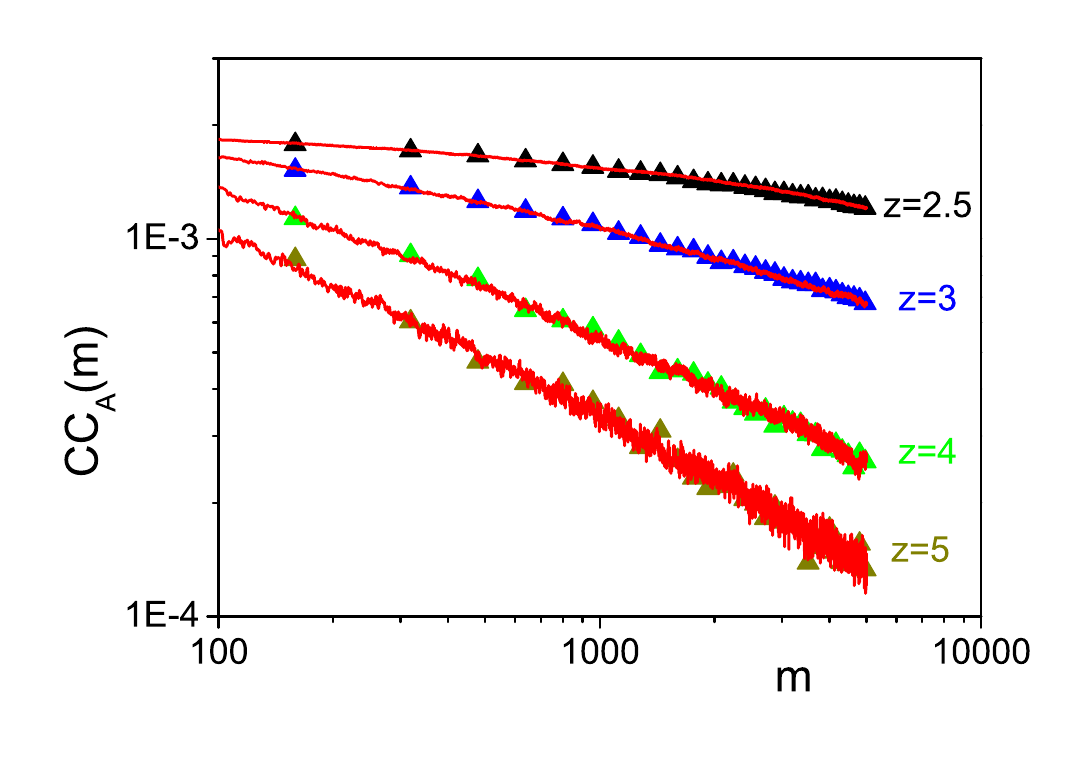}
\caption{\label{fig3} (color online). The cross-correlation function $CC_A(m)$ for the symbolic dynamics generated from the intermittent map of Eq.~(\ref{eq:eq1}) ($A=s$, triangles) and for the stochastic process defined in the text ($A=r$, red lines) for four different values of $z$: $z=2.5$ (black), $z=3$ (blue), $z=4$ (green), $z=5$ (dark yellow). For the numerical simulations we used an ensemble $10^5$ trajectories each of length $10^6$. }
\end{figure}

Based on the stochastic model introduced previously one can develop a method to find an analytical estimation $CC_a(m)$ of the cross-correlation function $CC_r(m)$. Such an attempt will enlighten further the origin of the found cross-correlations. Let us first consider two binary sequences $\{ x^{(i)} \}=\{ x^{(i)}_1, x^{(i)}_2, .., x^{(i)}_k,..\}$ and $\{ x^{(j)} \}=\{ x^{(j)}_1, x^{(j)}_2, .., x^{(j)}_n,..\}$ generated by the stochastic model. The function $CC_a(m)$ should be proportional to the joint probability $P_{ij}(x^{(i)}_{k}=1  ; x^{(j)}_{k+m}=1)$ that the random variable $x^{(i)}$ has the value $1$ at time step $k$ and the random variable $x^{(j)}$ has the value $1$ at time step $k+m$, averaged over the time:
\begin{equation}
CC_a(m) = \frac{1}{N-m} \sum_{k=1}^{N-m} P_{ij}(x^{(i)}_{k}=1 ; x^{(j)}_{k+m}=1)
\label{eq:eq3}
\end{equation}
Obviously it holds:
\begin{equation}
P_{ij}(x^{(i)}_{k}=1 ; x^{(j)}_{k+m}=1) = P(x^{(i)}_{k}=1) \cdot P(x^{(j)}_{k+m}=1)
\label{eq:eq4}
\end{equation}
since $x^{(i)}$ and $x^{(j)}$ are statistically independent. To calculate $P(x^{(i)}_{k}=1)$ one can use the method introduced in \cite{Proc1983} writing:
\begin{equation}
P(x^{(i)}_{k}=1)=\sum_{n=1}^{k-1} P(x^{(i)}_{k-n}=1) \cdot P_{1 \vert 1}(n \vert x^{(i)}_{k-n}=1) 
\label{eq:eq5}
\end{equation}
where $P_{1 \vert 1}(n \vert x^{(i)}_{k-n}=1)$ is the conditional probability to have a laminar phase of length $n$ directly after the instant $k-n$ if $x^{(i)}$ has the value $1$ at the time instant $k-n$. The appearance of a laminar phase with duration $n$ is independent of the value of $x^{(i)}$ at the instant $k-n$. Thus we find:
\begin{equation}
P_{1 \vert 1}(n \vert x^{(i)}_{k-n}=1)=\rho(n)~~~~~;~~~~~~\rho(n) \sim n^{-\frac{z}{z-1}}~~~;~~~n \gg 1
\label{eq:eq6}
\end{equation}
where $\rho(n)$ is the laminar length distribution. Inserting Eq.~(\ref{eq:eq6}) into Eq.~(\ref{eq:eq5}) we obtain the equation
\begin{equation}
P(x^{(i)}_{k}=1)=\sum_{n=1}^{k-1} P(x^{(i)}_{k-n}=1) \cdot \rho(n) 
\label{eq:eq7}
\end{equation}
which can be solved recursively using as initial condition $P(x^{(i)}_1=1)=p_0$ with $p_0 \in (0,1)$. A similar equation is obtained also for $P(x^{(j)}_{k+m}=1)$ replacing simply $k$ with $k+m$. Having solved Eq.~(\ref{eq:eq7}) one can then calculate the sum in Eq.~(\ref{eq:eq3}) to obtain an analytical expression  for $CC_a(m)$ containing three sums. The validity of the introduced analytical scheme is tested in Fig.~4 where we show the symbolic dynamics result $CC_s(m)$ together with the analytical form $CC_a(m)$ for $z=3$. We observe a very good agreement between the analytical result and the numerical simulations for $CC_s(m)$. Notice that $CC_r(m)$ is not shown in this plot. However, as illustrated in Fig.~3 the results for $CC_r(m)$ and $CC_s(m)$ are very close to each other for any considered $z$ and therefore $CC_a(m)$ can be used as an analytical estimation of the cross-correlation $CC_s(m)$ too. The analytical treatment leads us to the conclusion that it is the long-range character of the correlation between $P(x^{(i)}_k=1)$ and $P(x^{(j)}_{k\prime}=1)$ existing for any pair of intermittent trajectories which generates the observed cross-correlations. Note that this property has been discussed in \cite{Izra2007} in a different context.

\begin{figure}[htbp]
\includegraphics[width=8.5 cm]{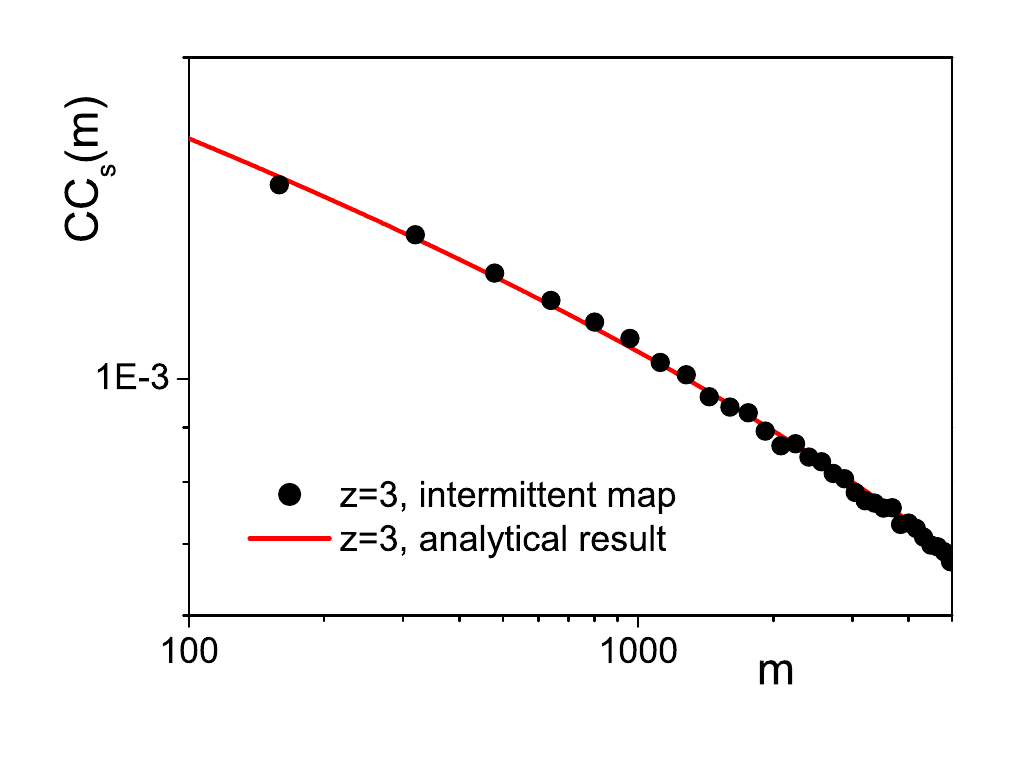}
\caption{\label{fig4} (color online). The cross-correlation functions $CC_s(m)$ (symbolic dynamics, intermittent map, black circles) and the analytical estimation $CC_a(m)$ (red line) for $z=3$.  }
\end{figure}

With our analysis we have demonstrated a mechanism to establish power-law cross-correlations between particles which do not interact with each other. This phenomenon is induced by the strong intermittent dynamics performed by each of the particles independently. Strong intermittency (sporadicity) discussed in this Letter is a result of the interaction of a particle with a suitable external potential (field)\footnote{This could be also a mean field generated by particle-particle interactions}. The appearance of long-ranged cross-correlations deems sporadic dynamics a plausible mechanism for the collective behaviour emerging in a $N$-particle system. Furthermore, since such a collective behaviour is accompanied by scale free inter-particle correlations, it could be related to the emergence of critical behaviour in the considered system. In fact a connection of intermittent dynamics with criticality has already been established in \cite{Conto2002} using the example of the $3-D$ Ising model. There it has been shown that the order parameter fluctuations at the critical point can be efficiently described by an intermittent map of Pomeau-Manneville type (similar to that of Eq.~(\ref{eq:eq1})) with additive noise. The exponent $z$ in this intermittent map is related to the isothermal critical exponent $\delta$ associated with the second order transition. This property sets a bound $z \geq 2$ necessary for the occurrence of critical behaviour. It is remarkable that this bound coincides with the bound obtained by our present analysis in order to have a divergent mean laminar length. An astonishing feature of our results is that the power-law cross-correlations emerge even without interaction among the particles. In the context of critical phenomena such a property is welcome since it could explain universality aspects. Indeed the microscopic interactions between the elementary degrees of freedom of a critical system do not play any role for the determination of the critical exponents and the associated scaling laws describing the phenomenology of an extended system at the critical point.

In the framework of our approach the obtained correlations are determined by the time evolution of the trajectories of two different particles. To enable a closer relation to equilibrium critical phenomena one should extend these ideas also to the case of a field depending both on time and on space. Such an extension requires the use of matrix equations for the field evolution replacing the variable $x^{(i)}_n$ by a scalar field $\phi(i,n)$ where $i$ is a spatial variable while $n$ is the time variable. At a first glance one could argue that for the calculation of the spatial cross-correlations one might exchange the role of spatial and temporal variables in the dynamics, use Eq.~(\ref{eq:eq1}) to describe changes of the field $\phi$ in space and average over the time variable. This would lead to power-law cross-correlations between the field values at different locations which is typical for a critical system. However a consistent treatment of this case requires more elaborate and extensive studies left for future investigations.
  
\begin{acknowledgments}
This work was made possible by the facilities of the Shared Hierarchical Academic Research Computing Network (SHARCNET:www.sharcnet.ca) and Compute/Calcul Canada. The authors thank C. Petri and B. Liebchen for fruitful discussions. The performed research has been co-financed by the European Union (European Social Fund --- ESF) and Greek national funds through the Operational Program "Education and Lifelong Learning" of the National Strategic Reference Framework (NSRF) - Research Funding Program: Heracleitus II. Investing in knowledge society through the European Social Fund. The authors also thank the IKY and DAAD for financial support in the framework of an exchange program between Greece and Germany (IKYDA 2010) and the UK ESPCI for funding under grant EP/E501311/1.
\end{acknowledgments}

\end{document}